\documentclass[twocolumn]{revtex4}
\usepackage{graphicx}
\usepackage{dcolumn}
\usepackage{amsmath}

\begin{document}

\title{Nambu-Goldstone-Leggett modes in multi-condensate superconductors}

\author{Takashi Yanagisawa}

\affiliation{Electronics and Photonics Research Institute,
National Institute of Advanced Industrial Science and Technology,
1-1-1 Umezono, Tsukuba, Ibaraki 305-8568, Japan}

\begin{abstract}
Multi-gap superconductors exhibit interesting properties.
In an $N$-gap superconductor, we have in general $U(1)^N$ phase
invariance. This multiple-phase invariance is partially or totally
spontaneously broken in a superconductor.
The Nambu-Goldstone modes, as well as Higgs modes, are important and will 
play an important role in multi-condensate superconductors. 
The additional phase invariance leads to a new quantum phase, with help of 
frustrated Josephson effects, 
such as the time-reversal symmetry breaking, the emergence of massless 
modes and fractionally quantized-flux vortices.
There is a possibility that half-flux vortices exist in two-component 
superconductors in a magnetic field. 
The half-quantum flux vortex can be interpreted as a monopole, and 
two half-flux vortices form a bound state connected by a 
domain wall.
There is an interesting analogy between quarks and fractionally 
quantized-flux vortices in superconductors.  
\end{abstract}



\maketitle

\section{Introduction} 

The study of multi-band superconductors started from works by
Moskalenko\cite{mos59}, 
Suhl et al.\cite{suh59}, Peretti\cite{per62} and Kondo\cite{kon63}, 
as a generalization
of the Bardeen-Cooper-Schrieffer (BCS) theory\cite{bcs57} to a multi-gap
superconductor.
Kondo first pointed out that the sign of gap function depends on the sign
of the pair-transfer interaction between two bands,
and the signs of two gaps are opposite
to each other when the pair-transfer interaction is repulsive.
The first observed two-band superconductor is Nb doped 
SrTiO$_3$\cite{bin80,gor15}.
The critical field $H_{c2}(0)$ and the sizable positive curvature of 
$H_{c2}(T)$ in
YNi$_2$B$_2$C and LuNi$_2$B$_2$C were analyzed within an effective
two-band model on the basis of multi-band Eliashberg theory\cite{shu98}.
Later, well known MgB$_2$\cite{nag01} and iron-based 
superconductors\cite{kam08} were discovered.
 
There are many interesting properties in multi-condensate
superconductors.
We show important characteristics in
the following:\\
(1) Multi-band superconductors have a possibility to exhibit high 
critical temperature $T_c$.
$T_c$ is always enhanced in the presence of interband interactions for
$s$-wave superconductors. 
MgB$_2$\cite{nag01} and iron-based
superconductors\cite{kam08} are multi-band superconductors with
relatively high $T_c$.  We also mention that 
layered cuprates\cite{ben04, bia13, bia87} can be regarded as
multi-gap superconductors.\\
(2) Unusual isotope effect has been observed in multi-band
superconductors.  This depends on the nature of the attractive
interaction in the pairing mechanism\cite{bus11,cho09,shi09,yan09} 
The isotope exponent $\alpha$ of (Ba,K)Fe$_2$As$_2$ takes values even in the
range of $\alpha<0$ and $\alpha>0.5$, depending on the property of
glue, especially strength and
the range of attractive interactions\cite{shi09,yan09}.\\
(3) In $N$-gap superconductors, the gap functions are written as
$\Delta_j=|\Delta_j|e^{i\theta_j}$ for $j=1,\cdots,N$. The $U(1)^N$
phase invariance at most can be
spontaneously broken.  The Coulomb repulsive interaction turns
the one-phase mode $\Phi=c_1\theta_1+\cdots+c_N\theta_N$ into a gapped 
plasma mode.
Thus there are at most $N-1$ modes and they can be low-energy excitation
modes in superconductors.  
These modes are in general massive due to Josephson interactions.
There is, however, a possibility that some of these modes become 
massless Nambu-Goldstone modes when the Josephson couplings are
frustrated.
The Josephson couplings between different bands will bring about
attractive phenomena; they are (a) time-reversal symmetry breaking
(TRSB)\cite{sta10,tan10a,tan10b,dia11,yan12,hu12,sta12,pla12,mai13,wil13,
tak14,yer15,gan14},
(b) the existence of massless (gapless) 
modes\cite{yan13,lin12,kob13,koy14,yan14,tan15,lin12b}
and low-lying excited states,
and (c) the existence of kinks and fractionally-quantized-flux
vortices\cite{izu90,vol09,tan02,bab02,kup11}.
The phase-difference mode between two gaps is sometimes called the Leggett
mode\cite{leg66}.
This mode will yield new excitation
modes in multi-gap superconductors.
The Leggett mode is realized as a Josephson plasma oscillation in
layered superconductors.
\\
(4) The existence of fractionally quantized-flux vortices is very
significant and attractive.  The kink (soliton) solution of phase
difference leads to a new mode and the existence of
half-quantum flux vortices in two-gap superconductors.
A generalization to a three-gap superconductor is not trivial and results 
in very attractive features, that is,  chiral states with time-reversal
symmetry breaking and the existence of fractionally quantized
vortices\cite{sta10,tan10a,tan10b,yan12}.
Further, in the case with more than four gaps, a new state is
predicted with a gapless excitation mode\cite{tan11}.
\\
(5) A new type of superconductors, called the 1.5 type as an
intermediate of types I and II, has been proposed for two-gap
superconductors\cite{mos09,sil11}.  This state may be realized as a result
of a multi-band
effect, and does not occur in a single-band superconductor.
\\
(6)
There is an interesting and profound analogy between particles physics
and superconductivity.  For example,
there is a similarity between the Dirac equation and the gap
equation of superconductivity\cite{nam60,nam97}.

In this paper we give a short review on several interesting properties 
concerning
Nambu-Goldstone modes in multi-condensate superconductors.
We focus on superconductors in the clean limit, and impurity effects
in multi-band superconductors are left for future studies.
The paper is organized as follows.
In section II we give a brief survey on multi-band superconductivity.
We examine the effective action and discuss the
plasma and Leggett modes in section III. 
We discuss the Higgs mode briefly in section IV.
Section V is devoted to a discussion on time-reversal symmetry
breaking.
In section VI we show that the half-quantized-flux vortex can be
regarded as a monopole in a multi-gap superconductor.
In section VII we discuss the emergency of massless Nambu-Goldstone
mode when there is a frustration between Josephson couplings.
We investigate a ${\bf Z}_2$-symmetry breaking where fluctuations
restore time reversal symmetry from the ground state with time-reversal
symmetry breaking in the subsequent section VIII.
In section IX we examine an $SU(N)$ sine-Gordon model.
This model is a generalization of the sine-Gordon model to that with
multiple variables and is regarded as a model of $G$-valued fields
for a Lie group $G$. 
This model is reduced to a 
unitary matrix model in some limit.
We give a summary in the last section.

\section{A brief survey}

Two years after the BCS theory was proposed\cite{bcs57}, an extension to
two overlapping bands was considered by Moskalenko\cite{mos59} and
Suhl, Matthias and Walker\cite{suh59}.  After these works, 
Peretti\cite{per62}, Kondo\cite{kon63} and Geilikman\cite{gei67}
reconsidered superconductors with multiple bands.
The motivation of Kondo's work is to understand the small isotope
effect observed for some transition metal superconductors.
Kondo investigated the exchange-like integral between different
bands, which is a non-phonon effective attractive interaction,
and proposed a possibility of small, being less than 0.5, or 
vanishing of the isotope effect of the critical temperature $T_c$
using the two-band model.
It was found by early works that the critical temperature is enhanced
higher than both of critical temperatures of uncoupled superconductors
due to the interband coupling.
The Ginzburg-Landau model was extended to include two conduction
bands\cite{yan12,til64,iva09,orl13}.
Kondo, at the same time, introduced different phases
assigned to two different gaps with phase difference $\pi$.
This indicates that we can take the phase difference  $\varphi$ to be 0 
or $\pi$ for
the two-band model.
A simple generalization to a three-band model was investigated much
later than Kondo's work.  It was shown independently\cite{sta10,tan10a,tan10b}
that the phase difference other than 0 or $\pi$ is possible.
It was indicated that the intermediate value of the phase difference $\varphi$
leads to time reversal symmetry breaking, which is a new state in
three-band superconductors.
There have been many works for a pairing state with time reversal symmetry
breaking\cite{dia11,yan12,hu12,sta12,pla12,mai13,wil13,tak14,yan13,lin12,
lin12b,hua14,yer14,tan15b} with relation to iron-base 
superconductors\cite{sto13}, and also from the viewpoint of
holographic superconductors\cite{cai13,wen14,nis15}. 

Leggett\cite{leg66} considered small fluctuation of phase difference,
which yields fluctuation in the density of Cooper pairs. 
This indicates a possibility of a collective excitation of phase
difference mode.
Leggett examined the Josephson term $-J\cos(\varphi)$ perturbatively
using  $\cos(\varphi)=1-(1/2)\varphi^2+\cdots$.
In the presence of large fluctuation of $\varphi$ we are not allowed
to use this approximation.
In this situation we must employ a sine-Gordon model.
This model has a kink solution\cite{raja} with fluctuation from
$\varphi=0$ to $2\pi$, which results in a new collective 
mode\cite{tan02,tan01,bab02b,cho06,eto14,fio14}.

An intensive study of multi-gap superconductivity started since
the discovery of MgB$_2$, and especially iron-based superconductors.
A new kind of superconductivity, called the type 1.5, was proposed
for MgB$_2$\cite{mos09} where it seems that there is an attractive
inter-vortex interaction preventing the formation of Abrikosov
vortex lattice. 
A theoretical prediction was given based on the model with
vanishing Josephson coupling\cite{bab05}.
There are some controversial on this subject\cite{kog11,bab12,kog12}.
We expect that the Higgs mode plays a role in this issue because
Higgs mode will produce an attractive force between vortices.
A three-band model is now considered as a model for iron-based
superconductors and the time reversal symmetry breaking is
investigated intensively.

\section{Plasma and Leggett modes}

Let us consider the Hamiltonian for multi-gap superconductors:
\begin{eqnarray}
H&=& \sum_{i\sigma}\int d{\bf r}\psi_{i\sigma}^{\dag}({\bf r})K_i({\bf r})
\psi_{i\sigma}({\bf r})\nonumber\\
&-& \sum_{ij}g_{ij}\int d{\bf r}\psi_{i\uparrow}^{\dag}({\bf r})
\psi_{i\downarrow}^{\dag}({\bf r})\psi_{j\downarrow}({\bf r})
\psi_{j\uparrow}({\bf r}),
\end{eqnarray}
where $i$ and $j$ (=1,2,$\cdots$)  are band indices.
$K_i({\bf r})$ stands for the kinetic operator:
$K_i({\bf r})= p^2/(2m_i)-\mu\equiv\xi_i({\bf p})$ where $\mu$ is the
chemical potential.
We assume that $g_{ij}=g_{ji}^*$.
The second term indicates the pairing interaction with the coupling 
constants $g_{ij}$.
This model is a simplified version of multi-band model where
the coupling constants $g_{ij}$ are assumed to be constants. 

In the functional-integral formulation, using the Hubbard-Stratonovich
transformation,
the partition function is expressed as follows:
\begin{eqnarray}
Z&=& \int d\psi_{\uparrow}d\psi_{\downarrow}\int d\Delta^*d\Delta\nonumber\\
&\times& \exp\left( -\int_0^{\beta}d\tau d^dx\sum_{ij}\Delta_i^*(G^{-1})_{ij}\Delta_j\right)
\nonumber\\
&\times& \exp\left( -\sum_j\int d\tau d^dx
(\psi_{j\uparrow}^{*}\psi_{j\downarrow})D
\left(
\begin{array}{c}
\psi_{j\uparrow} \\
\psi_{j\downarrow}^* \\
\end{array}
\right)
\right), 
\end{eqnarray}
where $G=(g_{ij})$ is the matrix of coupling constants and
\begin{eqnarray*}
D=\left(
\begin{array}{cc}
\partial_{\tau}+\xi_j({\bf p}) & \Delta_j \\
\Delta_j^* & \partial_{\tau}-\xi_j({\bf p}) \\
\end{array}
\right).
\end{eqnarray*}

$(G^{-1})_{ij}$ ($i\neq j$) indicates the Josephson coupling.
The condition for the matrix $G$ has been discussed in Ref.\cite{mar13}.

In order to obtain the effective action for phase variables $\theta_j$,
we perform the gauge transformation
\begin{eqnarray}
\left(
\begin{array}{c}
\psi_{j\uparrow} \\
\psi_{j\downarrow}^* \\
\end{array}
\right)\rightarrow \left(
\begin{array}{c}
e^{i\theta_j}\psi_{j\uparrow} \\
e^{-i\theta_j}\psi_{j\downarrow}^* \\
\end{array}
\right),
\end{eqnarray}
so that $\Delta_j$ are real and positive.
The effective action is written in the form
\begin{equation}
S= \sum_{ij}\int d\tau d^dx\Delta_i(G^{-1})_{ij}\Delta_j
\cos(2(\theta_i-\theta_j))
- {\rm Tr}\ln\tilde{D}\\
\end{equation}
where $\tilde{D}$ is given by the following matrix
\begin{eqnarray*}
\left(
\begin{array}{cc}
\partial_{\tau}+i\partial_{\tau}\theta_j+\xi_j({\bf p}+\nabla\theta_j) & \Delta_j \\
\Delta_j & \partial_{\tau}+i\partial_{\tau}\theta_j-\xi_j({\bf p}+\nabla\theta_j) \\
\end{array}
\right).
\nonumber\\
\end{eqnarray*}
We define the fluctuation mode (Higgs mode) $\eta_j$ of the amplitude of 
$\Delta_j$ as
\begin{equation}
\Delta_j= \Delta_{0j}+\eta_j,
\end{equation}
where $\Delta_{0j}$ is the gap function given by
the saddle point approximation.

The effective action for phase modes $\theta_j$ is given by the
usual quadratic form with the Josephson coupling.
For the two-band model with equivalent bands, i.e., $\xi_1=\xi_2=p^2/(2m)-\mu$
for simplicity, the Lagrangian reads
\begin{eqnarray}
L&=& \rho_F(\partial_{\tau}\theta_1-e\Phi)^2
+\rho_F(\partial_{\tau}\theta_2-e\Phi)^2 \nonumber\\
&&+\frac{n_s}{2m}\left( (\nabla\theta_1)^2+(\nabla\theta_2)^2 \right)
+\frac{1}{8\pi}(\nabla\Phi)^2 \nonumber\\
&&+ 2\gamma\Delta_0^2\cos(2(\theta_1-\theta_2))+\cdots
\nonumber\\
&=& 2\rho_F\left( \frac{1}{2}\partial_{\tau}\phi-e\Phi\right)^2
+\frac{n_s}{4m}(\nabla\phi)^2+\frac{1}{8\pi}(\nabla\Phi)^2
\nonumber\\
&&+\frac{1}{2}\rho_F(\partial_{\tau}\varphi)^2
+\frac{n_s}{4m}(\nabla\varphi)^2+2\gamma\Delta_0^2\cos(2\varphi)
+\cdots ,
\end{eqnarray}
where $\cdots$ indicates higher order terms including the coupling terms  
between amplitude modes and phase modes.
We introduced the scalar potential $\Phi$ which represents the
Coulomb interaction and defined
\begin{equation}
\phi= \theta_1+\theta_2,~~~ \varphi=\theta_1-\theta_2.
\end{equation}
$\gamma$ denotes the Josephson coupling strength given by
$\gamma=\gamma_{12}\equiv (G^{-1})_{12}$.
$\rho_F$ is the density of states at the Fermi level and $n_s$ is the
electron density per band given by 
$n_s=n_j\equiv (1/V)\sum_k[1-(\xi_j(k)/E_j)(1-2f(E_j)]$.
The derivative of the total phase $\nabla\phi$ represents the plasma mode 
with the plasma
frequency $\omega_p^2=4\pi n_{e}e^2/m$ where $n_e$ is the total electron 
density including up and down spins.
This is seen by writing the terms of $\phi$ in the following form by integrating
out the scalar potential $\Phi$:
\begin{equation}
\frac{1}{2}\rho_F\left( \frac{\omega_n^2}{{\bf k}^2+16\pi\rho_Fe^2}
+ \frac{n_s}{2m\rho_F}\right)
{\bf k}^2|\phi(i\omega_n,{\bf k})|^2,
\end{equation}
after the Fourier transformation.  
This indicates that the plasma mode is described by the derivative of the
total phase $\nabla\phi$.
By performing the analytic continuation $i\omega_n\rightarrow \omega+i\delta$,
we obtain the dispersion relation as
\begin{equation}
\omega^2 = \omega_{pl}^2+c_s^2{\bf k}^2,
\end{equation}
where $\omega_{pl}^2=8\pi n_se^2/m$ and
$c_s=v_F/\sqrt{3}$ with the Fermi velocity $v_F$.
We have $n_s=n_e/2$ and $\omega_{pl}^2=\omega_p^2$ at absolute zero
$T=0$.

The Lagrangian of the phase difference mode (Leggett mode) $\varphi$ is
given by the sine-Gordon model.
This mode is a massless mode if the
Josephson coupling $\gamma$ vanishes.
When $\varphi$ is small, the sine-Gordon model describes an oscillation
mode, by using $\cos\varphi=1-\varphi^2/2+\cdots$.
We assume that $\gamma$ is positive so that $\varphi$ describes a
stable oscillation mode.
The frequency of this mode is proportional to
the gap function:
\begin{equation}
\omega_J= 2\sqrt{\frac{2\gamma}{\rho_F}}\Delta_0.
\end{equation}
The dispersion relation is given as
\begin{equation}
\omega^2 = \omega_J^2+\frac{1}{3}v_F^2{\bf k}^2.
\end{equation}
This kind of oscillation mode is known as the Josephson plasma
mode\cite{kle92, ohy92, tam92, mat95, koy96}.
In MgB$_2$ the frequency of the oscillation mode (Leggett mode) was
estimated to be 1.6 or 2THz\cite{sha02}.
There are two superconducting gaps in MgB$_2$; their magnitudes are given 
by $\Delta_1\simeq 1.2{\rm meV}-3.7{\rm meV}$ ($\pi$ band, smaller gap) 
and $\Delta_2\simeq 6.4{\rm meV}-6.8{\rm meV}$
($\sigma$ band, larger gap)\cite{cho02}.
Thus, the frequency of the Leggett mode is larger than $2\Delta_1$.
The observation of the Leggett mode in MgB$_2$ was recently reported by
Raman scattering measurements\cite{blu07}.

The effective action density for Nambu-Goldstone modes for an $N$-band
superconductor is written as
\begin{eqnarray}
L_{\theta}&=& \sum_j\rho_j(\partial_{\tau}\theta_j-e\Phi)^2
+\sum_j\frac{n_{j}}{2m_j}(\nabla\theta_j)^2 \nonumber\\
&&+\sum_{j\ell}\Delta_j(G^{-1})_{j\ell}\Delta_{\ell}\cos(2(\theta_j-\theta_{\ell}))
+\frac{1}{8\pi}(\nabla\Phi)^2,\nonumber\\
\end{eqnarray}
where $\rho_j$, $n_{j}$ and $m_j$ are the density of states, the
electron density and the electron mass in the $j$-th band, respectively.
For $N=2$, this action density reads
\begin{eqnarray}
L_{\theta}&=& \rho_1(\partial_{\tau}\theta_1-e\Phi)^2
+\rho_2(\partial_{\tau}\theta_2-e\Phi)^2 \nonumber\\
&&+\frac{n_{1}}{2m_1}(\nabla\theta_1)^2+\frac{n_{2}}{2m_2}(\nabla\theta_2)^2
 \nonumber\\
&&+ 2\gamma_{12}\Delta_1\Delta_2\cos(2(\theta_1-\theta_2))
+\frac{1}{8\pi}(\nabla\Phi)^2. 
\end{eqnarray}
When we neglect the scalar potential $\Phi$ for neutral superconductors, 
the dispersion relations of the Nambu-Goldstone mode and the Leggett mode are
determined by the condition that the determinant of the $2\times 2$ matrix
vanishes\cite{sha02,ota11}:
\begin{eqnarray*}
\left(
\begin{array}{cc}
\rho_1\omega^2-\frac{n_{1}}{2m_1}k^2-4\gamma\Delta_{01}\Delta_{02} &
4\gamma\Delta_{01}\Delta_{02} \\
4\gamma\Delta_{01}\Delta_{02} &
\rho_2\omega^2-\frac{n_{2}}{2m_2}k^2-4\gamma\Delta_{01}\Delta_{02} \\
\end{array}
\right).\nonumber\\
\end{eqnarray*}
The dispersion relations of the Nambu-Goldstone and Leggett modes are,
respectively, given by
\begin{eqnarray}
\omega^2 &=& \frac{1}{\rho_1+\rho_2}\left( \frac{n_{1}}{2m_1}
+\frac{n_{2}}{2m_2} \right)k^2 = v_N^2k^2,\\
\omega^2 &=& 4\frac{\rho_1+\rho_2}{\rho_1\rho_2}\gamma\Delta_{01}\Delta_{02}
+ \frac{1}{\rho_1+\rho_2}\left( \frac{n_{1}\rho_2}{2m_1\rho_1}
+\frac{n_{2}\rho_1}{2m_2\rho_2} \right)k^2\nonumber\\
&=& \omega_J^2+v_L^2k^2,
\end{eqnarray}
where
\begin{eqnarray}
v_N^2 &=& \frac{1}{3}\frac{\rho_1v_{F1}^2+\rho_2v_{F2}^2}{\rho_1+\rho_2},\\
v_L^2 &=& \frac{1}{3}\frac{\rho_2v_{F1}^2+\rho_1v_{F2}^2}{\rho_1+\rho_2},\\
\omega_J^2 &=& 4\frac{\rho_1+\rho_2}{\rho_1\rho_2}\gamma\Delta_{01}\Delta_{02}.
\end{eqnarray}
$v_{Fj}$ is the Fermi velocity of the $j$-th band.

For charged superconductors, we must have the scalar potential $\Phi$ in $L$.
We integrate out the scalar potential $\Phi$ to obtain the effective action given
as after the Fourier transformation: 
\begin{eqnarray}
L_{\theta} &=& \frac{8\pi e^2\rho_1\rho_2}{k^2+8\pi e^2(\rho_1+\rho_2)}
(\partial_{\tau}\varphi)^2-4\gamma\Delta_1\Delta_2\varphi^2\nonumber\\ 
&&+\frac{n_{1}}{2m_1}(\nabla\theta_1)^2+\frac{n_{2}}{2m_2}(\nabla\theta_2)^2.
\nonumber\\
&& + \frac{k^2}{k^2+8\pi e^2(\rho_1+\rho_2)}\left( 
\rho_1(\partial_{\tau}\theta_1)^2+\rho_2(\partial_{\tau}\theta_2)^2 \right),
\nonumber\\
\end{eqnarray}
for $\varphi=\theta_1-\theta_2$.
In the long-wavelength limit ${\bf k}\rightarrow 0$, the quadratic
terms of the mode $\varphi$ are given by 
$\rho_1\rho_2/(\rho_1+\rho_2)\cdot(\partial_{\tau}\varphi)^2-4\gamma_{12}
\Delta_{01}\Delta_{02}\varphi^2$.
The dispersion of the Leggett mode is given by
\begin{eqnarray}
\omega^2 &=& \omega_J^2+ \frac{\rho_1+\rho_2}{\rho_1\rho_2}
\frac{1}{n_{1}/2m_1+n_{2}/2m_2}\frac{n_{1}}{2m_1}\frac{n_{2}}{2m_2}k^2
\nonumber\\
&=& \omega_J^2+\frac{1}{9}\frac{1}{v_N^2}v_{F1}^2v_{F2}^2k^2.
\end{eqnarray}
The dispersion of the plasma mode is
\begin{equation}
\omega^2 = \omega_{pl}^2+v_N^2k^2,
\end{equation}
where $\omega_{pl}$ is the plasma frequency for the two-band model.

In general, in an N-gap superconductor, there are $N-1$ Leggett modes
because one mode becomes a massive mode with the plasma frequency by coupling
to the Coulomb potential.
When N bands are equivalent, the Josephson term is invariant under
an $S_{N}$ group action.
When there is an anisotropy that breaks the equivalence among
several bands, we have lower symmetry than $S_{N}$.

\section{Higgs mode}

Let us discuss the fluctuation of the amplitude of gap functions, which
can be called the Higgs mode.
Recently, there has been an increasing interest in a role of the Higgs 
mode in superconductors\cite{mat13,mea14,bar13,koy14b,tsu15}.
The quadratic form of the action in the field $\eta_j$ is given as
\begin{eqnarray}
L_H &=& \sum_{j\ell}\eta_j(G^{-1})_{j\ell}\eta_{\ell}
\cos(2(\bar{\theta}_j-\bar{\theta}_{\ell})) \nonumber\\
&&+\frac{1}{2}\sum_{\ell}{\rm tr}G_{\ell}^{(0)}\sigma_1\eta_{\ell}
G_{\ell}^{(0)}\sigma_1\eta_{\ell},
\end{eqnarray}
where $\sigma_1$ is the Pauli matrix and $G_{\ell}^{(0)}$ is the
Green function defined by
\begin{eqnarray}
\left(G_{\ell}^{(0)}\right)^{-1}(i\omega_n,{\bf p})=\left(
\begin{array}{cc}
-i\omega_n+\xi({\bf p}) & \Delta_{\ell 0} \\
\Delta_{\ell 0} & -i\omega_n-\xi({\bf p}) \\
\end{array}
\right).\nonumber\\
\end{eqnarray}
$\bar{\theta}_j$ is a mean-field value of $\theta_j$, which was put
zero in the previous section.
The excitation spectra is determined from the condition
${\rm det}H=0$ where $H$ is the matrix $H=(H_{j\ell})$
defined by 
\begin{equation}
L_H=\sum_{j\ell}\eta_jH_{j\ell}\eta_{\ell}.
\end{equation}
This action is reduced to the Higgs part of Ginzburg-Landau functional when
the temperature $T$ is near $T_c$ with vanishing gap at $T=T_c$.
At low temperature the spectrum has a gap being proportional to
the mean-field gap function.
An effect of the Higgs mode in a superconductor has not been
sufficiently clarified yet and there may be a need for further
studies.

For $N=2$ (two-band superconductor), the matrix $(H_{j\ell})$ is
written as
\begin{eqnarray}
\left(
\begin{array}{cc}
\gamma_{11}+\frac{1}{2}\Pi_1 & \gamma_{12} \\
\gamma_{21} & \gamma_{22}+\frac{1}{2}\Pi_2 \\
\end{array}
\right),
\end{eqnarray}
where $\Pi_{\ell}$ is 
\begin{eqnarray}
\Pi_{\ell}({\bf q},i\epsilon)&=& \frac{1}{\beta}\sum_n\frac{1}{V}
\sum_{{\bf p}}{\rm tr}\Big[G_{\ell}^{(0)}({\bf p}+{\bf q},i\omega_n+i\epsilon)
\sigma_1\nonumber\\
&&\times G_{\ell}^{(0)}({\bf p},i\omega_n)\sigma_1\big].
\end{eqnarray}
Then the dispersion relation of the Higgs mode $\omega=\omega({\bf q})$ 
is given by a 
solution of the equation
\begin{eqnarray}
1&+&\frac{1}{2}g_{11}\Pi_1({\bf q},\omega)
+\frac{1}{2}g_{22}\Pi_2({\bf q},\omega)\nonumber\\
&&+\frac{1}{4}{\rm det}G\cdot
\Pi_1({\bf q},\omega)
\Pi_2({\bf q},\omega)=0,
\end{eqnarray}
where ${\rm det}G=g_{11}g_{22}-g_{12}g_{21}$.

The first term in $L_H$ gives a correction to the Josephson coupling
when replacing $\eta_j\eta_{\ell}$ by the expectation value
$\langle\eta_j\eta_{\ell}\rangle$.  
A Higgs-Leggett coupling appears from the third term in $L_{\theta}$
in eq.(12).
This type of fluctuations will
result in some effect on the stability of unconventional states
which will be discussed in subsequent sections.
This subject concerning fluctuation effects is not, however, considered 
in this paper and is left
as a future problem.

\section{Time-Reversal Symmetry Breaking}

The gap function, defined as
$\Delta_i({\bf r})= -\sum_j g_{ij}\langle\psi_{j\downarrow}({\bf r})
\psi_{j\uparrow}({\bf r})\rangle$, satisfies the
gap equation
\begin{equation}
\Delta_i= \sum_j g_{ij}N_j\Delta_j\int d\xi_j\frac{1}{E_j}
\tanh\left(\frac{E_j}{2k_B T}\right),
\end{equation}
where $N_j$ is the density of states at the Fermi surface in the $j$-th band
and $E_j=\sqrt{\xi_j^2+|\Delta_j|^2}$.
$\Delta_i$ in this section is the mean-field solution in section III
which is obtained by a saddle-point approximation.
We set
\begin{equation}
\zeta_j= \int_0^{\omega_{Dj}} d\xi_j \frac{1}{E_j}\tanh
\left(\frac{E_j}{2k_B T}\right),
\end{equation}
and $\gamma_{ij}=(G^{-1})_{ij}$ where $G=(g_{ij})$.
We write the gap equation in the following form,
\begin{eqnarray}
\left(
\begin{array}{cccc}
\gamma_{11}-N_1\zeta_1  &  \gamma_{12}  &  \gamma_{13}  &  \cdots  \\
\gamma_{21}  &  \gamma_{22}-N_2\zeta_2  &  \gamma_{23}  &  \cdots \\
\gamma_{31}  &  \gamma_{32}  &  \gamma_{33}-N_3\zeta_3  &  \cdots  \\
\cdots  &  \cdots  &  \cdots  &  \cdots  \\
\end{array}
\right)\left(
\begin{array}{c}
\Delta_1  \\
\Delta_2  \\
\Delta_3  \\
\cdots  \\
\end{array}
\right)=0. \nonumber\\
\end{eqnarray}
$\gamma_{ij}$ ($i\neq j$) gives the interband Josephson coupling
between bands $i$ and $j$\cite{yan12}.

When the gap functions $\Delta_j$ are complex-valued functions,
the time-reversal symmetry is broken.  The condition for TRSB is
that the following equation for the imaginary part ${\rm Im}\Delta_j$
has a nontrivial solution:
\begin{eqnarray}
\left(
\begin{array}{ccc}
\gamma_{12}  &  \gamma_{13}  &  \cdots  \\
\gamma_{22}-N_2\zeta_2  &  \gamma_{23}  &  \cdots \\
\gamma_{32}  &  \gamma_{33}-N_3\zeta_3  &  \cdots  \\
\cdots  &  \cdots  &  \cdots  \\
\end{array}
\right)\left(
\begin{array}{c}
{\rm Im}\Delta_2  \\
{\rm Im}\Delta_3  \\
\cdots  \\
\end{array}
\right)=0,
\label{imgeq}
\end{eqnarray}
where we adopt that $\Delta_1$ is real for simplicity and $\gamma_{ij}$ are
real.  We assume that $\gamma_{ij}=\gamma_{ji}$.
In the case of $N=3$, the condition for TRSB has been obtained\cite{hu12,wil13}.
We have a necessary condition 
$\gamma_{12}\gamma_{23}\gamma_{13}>0$\cite{tan10a,tan10b}.
The determinant of each $2\times 2$ matrix in eq.(\ref{imgeq}) should 
vanish so that non-trivial solution ${\rm Im}\Delta_j$ ($j=2,3$) exist.
Then we have
\begin{eqnarray}
\gamma_{12}\gamma_{23}-(\gamma_{22}-N_2\zeta_2)\gamma_{13}&=& 0,\\
(\gamma_{22}-N_2\zeta_2)(\gamma_{33}-N_3\zeta_3)-\gamma_{23}^2&=& 0,\\
\gamma_{12}(\gamma_{33}-N_3\zeta_3)-\gamma_{12}\gamma_{23}&=& 0.
\end{eqnarray}
When we assume $\gamma_{13}\neq 0$, we obtain
\begin{equation}
\gamma_{22}-N_2\zeta_2= \gamma_{12}\gamma_{23}/\gamma_{13}.
\label{gamma2}
\end{equation}
Similarly, we have by assuming $\gamma_{12}\neq 0$
\begin{equation}
\gamma_{33}-N_3\zeta_3 = \gamma_{23}\gamma_{13}/\gamma_{12}.
\label{gamma3}
\end{equation}
From the gap equation 
$\gamma_{21}\Delta_1+(\gamma_{22}-N_2\zeta_2)\Delta_2+\gamma_{23}\Delta_3=0$,
we obtain the relation
\begin{equation}
\frac{\Delta_1}{\gamma_{23}}+\frac{\Delta_2}{\gamma_{31}}
+\frac{\Delta_3}{\gamma_{12}}=0.
\label{triangle}
\end{equation}
The complex numbers $\Delta_1/\gamma_{23}, \cdots$ form a triangle in
the TRSB state.
The transition form TRSB to the state with time-reversal symmetry
takes place when  the triangle relation is broken.
From eqs.(\ref{gamma2}) and (\ref{gamma3}), the critical temperature 
$T_c$ should satisfy
\begin{equation}
N_j\ln\left( \frac{2e^{\gamma_E}\omega_{Dj}}{\pi k_BT_c}\right)
= \gamma_{jj}-\frac{\gamma_{jn}\gamma_{jm}}{\gamma_{nm}},
\end{equation}
where $j$, $n$ and $m$ are different to one another and $\gamma_E$ is the
Euler constant.
The stability of TRSB state has been examined by evaluating the
free energy\cite{sta10,wil13,mar13}

In the simplest case where all the bands are equivalent and  $\gamma_{ij}$
($i\neq j$) are the same, the chiral state in Fig.1 is realized.
We have $(\theta_1,\theta_2,\theta_3)=(0,2\pi/3,4\pi/3)$ for Fig.1(a)
and $(\theta_1,\theta_2,\theta_3)=(0,4\pi/3,2\pi/3)$ for Fig.1(b).
The two states are degenerate and have chirality $\kappa=1$ and
$\kappa=-1$, respectively, where the chirality is defined by
$\kappa=(2/3\sqrt{3})[\sin(\theta_1-\theta_2)+\sin(\theta_2-\theta_3)
+\sin(\theta_3-\theta_1)]$.
In the chiral state $\Delta_1/\gamma_{23}, \cdots$ form an equilateral triangle.
In this case the eigenvalues of the gap equation are degenerate and the
chiral TRSB state is realized. 

For $N>3$ it is not straightforward to derive the condition for TRSB.
We consider here a separable form for the Josephson couplings:
\begin{equation}
\gamma_{ij}=\gamma_i\gamma_j~~{\rm for}~~i\neq j,
\end{equation}
where $\gamma_j$($\neq 0$) ($j=1,\cdots,N$) are real constants.
The condition $\gamma_{12}\gamma_{23}\gamma_{31}=\gamma_1^2\gamma_2^3\gamma_3^2>0$
is satisfied.
For $N=4$ we obtain from eq.(\ref{imgeq})
\begin{equation}
\frac{\Delta_1}{\gamma_2\gamma_3\gamma_4}
+\frac{\Delta_2}{\gamma_3\gamma_4\gamma_1}
+\frac{\Delta_3}{\gamma_4\gamma_1\gamma_2}
+\frac{\Delta_4}{\gamma_1\gamma_2\gamma_3}=0.
\end{equation}
Then the triangle condition in eq.(\ref{triangle}) is generalized to the polygon
condition
for general $N\ge 3$:
\begin{equation}
\frac{\Delta_1}{\gamma_2\gamma_3\cdots\gamma_N}
+\frac{\Delta_2}{\gamma_3\gamma_4\cdots\gamma_N\gamma_1}+\cdots
+\frac{\Delta_N}{\gamma_1\gamma_2\cdots\gamma_{N-1}}=0.
\label{polygon}
\end{equation}
We assume that the polygon is not crushed to a line, which means, in the
case $N=3$, the triangle inequality holds.
Under these conditions, the solution with time-reversal symmetry breaking
exists and massless excitation modes also exist at the same.
The existence of massless modes will be examined in next section.

\begin{figure}
\begin{center}
  \includegraphics[height=6cm,angle=90]{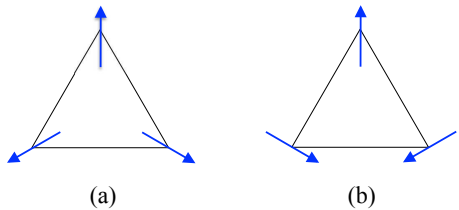}
\caption{Chiral state with time-reversal symmetry breaking.
Two states have the chirality $\kappa=+1$ for (a) and $\kappa=-1$ for (b).
}
\label{fig:1}       
\end{center}
\end{figure}

\section{Half quantum-flux vortex and a Monopole}

The sine-Gordon model has been studied to investigate a new
dynamics of multi-gap superconductors\cite{tan02,bab02,gur03}.
When the oscillation of phase difference $\varphi$ is small,
we can expand the potential around a minimum.  This results in
the Leggett mode as described in section III. 
In the presence of large oscillation, we cannot use a perturbative
method and we must consider a non-perturbative kink solution.
This leads to a half-quantum flux vortex.

The sine-Gordon model has a kink solution\cite{raja}.
If we impose the boundary condition such that $\varphi\rightarrow 0$ as
$x\rightarrow -\infty$ and $\varphi\rightarrow 2\pi$ as $x\rightarrow\infty$,
we have a kink solution like $\varphi= \pi+2\sin^{-1}(\tanh(\sqrt{\kappa}x))$
for a constant $\kappa$.
The phase difference $\varphi$ should be changed from 0 to $2\pi$ 
to across the kink.
This means that $\theta_1$ changes from 0 to $\pi$ and at the
same time $\theta_2$ changes from 0 to $-\pi$.
In this case, a half-quantum-flux vortex exists at the edge of the kink.
This is shown in Fig.2 where the half-quantum vortex is at the
edge of the cut (kink).
A net change of $\theta_1$ is $2\pi$ by a counterclockwise
encirclement of the vortex, and that of $\theta_2$ vanishes.  Then,
we have a half-quantum flux vortex.

\begin{figure}
\begin{center}
\includegraphics[width=3cm,angle=90]{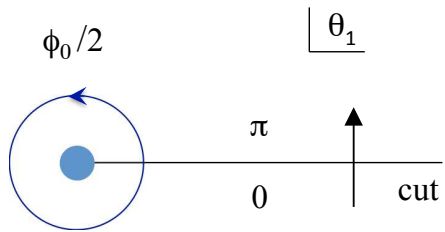}
\caption{
Half-quantum flux vortex with a line singularity (kink).
The phase variables $\theta_1$ changes from 0 to $\pi$ when
crossing a singularity.
}
\label{half-cut}
\end{center}
\end{figure}

The phase-difference gauge field ${\bf B}$ is defined as\cite{yan13}
\begin{equation}
{\bf B}=  -\frac{\hbar c}{2e^*}\nabla\varphi.
\end{equation}
The half-quantum vortex can be interpreted as a monopole\cite{yan12}.
Let us assume that there is a cut, namely, kink on the real axis
for $x>0$.  The phase $\theta_1$ is represented by
\begin{equation}
\theta_1 = -\frac{1}{2}{\rm Im}\log\zeta+\pi,
\end{equation}
where
$\zeta= x+iy$.
The singularity of $\theta_j$ can be transferred to a singularity of the
gauge field by a gauge transformation.
We consider the case $\theta_2=-\theta_1$: $\varphi=2\theta_1$.  Then we have
\begin{equation}
{\bf B}= -\frac{\hbar c}{2e^*}\nabla\varphi= -\frac{\hbar c}{e^*}\frac{1}{2}
\left( \frac{y}{x^2+y^2},-\frac{x}{x^2+y^2},0 \right).
\end{equation}
Thus, when the gauge field ${\bf B}$ has a monopole-type singularity,
the vortex with half-quantum flux exists in two-gap superconductors.

Let us consider the fictitious z axis perpendicular to the x-y plane.
The gauge potential (1-form) is given by
\begin{equation}
\Omega_{\pm}= -\frac{1}{2}\frac{1}{r(z\pm r)}(ydx-xdy)=\frac{1}{2}(\pm 1
-\cos\theta)d\phi,
\end{equation}
where $r=\sqrt{x^2+y^2+z^2}$, and $\theta$ and $\phi$ are Euler angles.
$\Omega_{\pm}$ correspond to the gauge potential in the upper and lower
hemisphere $H_{\pm}$, respectively.  $\Omega_{\pm}$ are connected by
$\Omega_+ = \Omega_-+d\phi$.
The components of $\Omega_+$ are
\begin{equation}
\Omega_{\mu}= \frac{1}{2}(1-\cos\theta)\partial_{\mu}\phi.
\end{equation}
At $z=0$, $\Omega_{\mu}$ coincides with the gauge field for half-quantum
vortex.  If we identify $\varphi$ with $\phi$, we obtain
\begin{equation}
{\bf B}= \frac{\hbar c}{e^*}{\bf \Omega},
\end{equation}
at $\theta=\pi/2$.
$\{\Omega_{\pm}\}$ is the $U(1)$ bundle $P$ over the sphere $S^2$.
The Chern class is defined as
\begin{equation}
c_1(P)= -\frac{1}{2\pi}F=-\frac{1}{2\pi}d\Omega_+.
\end{equation}
The Chern number is given as
\begin{eqnarray}
C_1&=& \int_{S^2}c_1 = -\frac{1}{2\pi}\int_{S^2}F \nonumber\\
&=& -\frac{1}{2\pi}\left( \int_{H_+}d\Omega_++\int_{H_-}d\Omega_-\right)=1.
\end{eqnarray}
In general, the gauge field ${\bf B}$ has the integer Chern number:
$C_1=n$.
For $n$ odd, we have a half-quantum flux vortex.

The half-flux vortex has been investigated in the study of $p$-wave
supercon- ductivity\cite{vol09,kee00,jan11}.
In the case of chiral $p$-wave superconductivity, the singularity
of $U(1)$ phase is, however, canceled by the kink structure of the
$d$-vector.  This is the difference between two-band superconductivity
and $p$-wave superconductivity.

As we can expect easily, a fractional quantum-flux vortex state is not
stable because the singularity (kink, domain wall) costs energy being
proportional to the square root of the Josephson coupling.
Thermodynamic stability was discussed in Ref.\cite{kup11}.
Two vortices form a molecule by two kinks.  This state may have lower
energy than the vortex state with a single quantum flux $\phi_0$
because the magnetic energy of two fractional vortices is smaller than
$\phi_0^2$ of the unit quantum flux.
The energy of kinks is proportional to the distance $R$ between two
fractional vortices when $R$ is large.  Thus, the attractive
interaction works between them when $R$ is sufficiently large.
There is an interesting analogy between quarks and fractional flux
vortices\cite{nit12}.

\section{Massless Nambu-Goldstone modes}

We examined the phase modes that are Nambu-Goldstone modes by nature
emerging due to a spontaneous symmetry breaking in section III. 
There, one mode becomes
massive by coupling to the scalar potential, called the plasma mode, 
and the other modes become massive due to Josephson couplings, called
the Leggett modes.
In this section we show that massive modes change into massless modes
when some conditions are satisfied.

The Josephson potential is given as
\begin{equation}
V \equiv -\sum_{i\neq j}\gamma_{ij}\Delta_{i0}\Delta_{j0}\cos(\theta_i-\theta_j),
\end{equation}
where $\gamma_{ij}=\gamma_{ji}$ are chosen real.
Obviously the phase difference modes $\theta_i-\theta_j$ acquire masses.
This would change qualitatively when $N$ is greater than 3 or equal to 3.
We discuss this in this section.

We show that massless modes exist for an $N$-equivalent frustrated
band superconductor.  Let us consider the potential for $N\ge 4$
given by
\begin{eqnarray}
V&=& \Gamma [ \cos(\theta_1-\theta_2)+\cos(\theta_1-\theta_3)+\cdots
+\cos(\theta_1-\theta_N)\nonumber\\
&+&\cdots+\cos(\theta_{N-1}-\theta_N) ].
\label{v-n}
\end{eqnarray}
For $\Gamma>0$, there are two massive modes and $N-3$ massless modes,
near the minimum
$(\theta_1,\theta_2,\theta_3,\theta_4,\cdots)=(0,2\pi/N,4\pi/N,6\pi/N,\cdots)$.
This can be seen by writing the potential in the form
\begin{equation}
V = \frac{\Gamma}{2}\Big[ \left(\sum_{i=1}^N{\bf S}_i\right)^2-N\Big],
\end{equation}
where ${\bf S}_i$ ($i=1,\cdots,N$) are two-component vectors with
unit length $|{\bf S}_i|=1$.
$V$ has a minimum $V_{min}=-\Gamma N/2$ for $\sum_i{\bf S}_i=0$.
Configurations under this condition have the same energy and
can be continuously mapped to each other with no excess energy.
At $(\theta_1,\theta_2,\cdots)=(0,2\pi/N,4\pi/N,\cdots)$ with
$V=-\Gamma N/2$,
satisfying $\sum_i{\bf S}_i=0$, the vectors ${\bf S}_i$ form a polygon.
The polygon can be deformed with the same energy (see Figs.3(a) and 3(b)).
The existence of massless modes was examined numerically for
the multi-gap BCS model\cite{kob13}.
It has been shown that there is a large region in the parameter space
where massless modes exist.

Let us discuss the Josephson potential in a separable form.
This is given by
\begin{equation}
V= \sum_{i\neq j}\gamma_{ij}\Delta_i^*\Delta_j
= \sum_{i\neq j}\gamma_i\gamma_j\Delta_i^*\Delta_j.
\end{equation}
This is written as
\begin{equation}
V= |P|^2-\sum_j\gamma_j^2|\Delta_j|^2,
\end{equation}
where $P=\sum_j\gamma_j\Delta_j$.
$V$ has a minimum when $P=0$ is satisfied.
$P=0$ is equivalent to the polygon condition in eq.(\ref{polygon}).
Because the polygon for $N>3$ can be deformed continuously without finite
excitation energy, a massless mode exists\cite{yan13} (Fig.3 (a) and (b)).
We have one massless mode for $N=4$ and two massless modes for $N=5$.
A spin model, corresponding to the Josephson model considered here, also
has gapless excitation modes.

When the polygon is crushed to a line, the time-reversal symmetry is not
broken.
A massless mode, however, exists when $P=0$.
An example is shown in Fig.3(c) called a linear model.
In this model there are two independent modes and
the quadratic term of one mode vanishes as can be shown by
explicit calculations.
A mode called the scissor mode becomes massless.

In this section we did not consider an effect of the amplitude
mode (Higgs mode) $\eta_j$.  This mode may be important when
discussing the stability of massless modes.
This is a future problem.

\begin{figure}
\begin{center}
\includegraphics[height=8cm,angle=90]{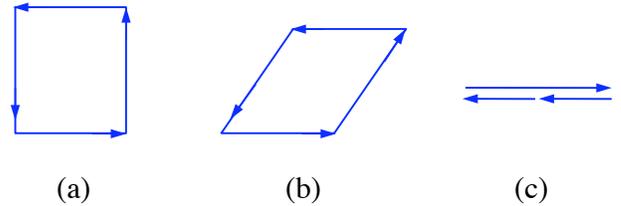}
\caption{Polygon state satisfying $\sum_j{\bf S}_j=0$ for $N=4$
in (a) and (b), where time-reversal symmetry is broken and a massless mode
exists.
A linear state with $\sum_j\gamma_j\Delta_j=0$ is shown in (c), where
a massless mode exists but the time-reversal symmetry is not broken.
}
\label{fig:2}       
\end{center}
\end{figure}

\section{${\bf Z}_2$ phase transition}


In this section we present a field theoretic model that shows a
chiral transition.  This model is extracted from a model for
multi-gap superconductors.
The model should be regarded as a model in field theory, and we also
discuss applicability to real superconductors.
We adopted the London approximation to derive the model, where the
fluctuation modes (Higgs modes) $\eta_j$ of the gap functions are 
neglected.
A role of the fluctuation mode concerning the existence of the phase
transition would be a problem for future discussion.

Let us consider an action for phase variable $\theta_j$:
\begin{eqnarray}
S[\theta]&=& \frac{1}{k_BT}\int d^dx\Big[
\sum_j\frac{n_{sj}}{2m_j}(\nabla\theta_j)^2 \nonumber\\
&+&\sum_{i\neq j}\gamma_{ij}\Delta_{i0}\Delta_{j0}\cos(\theta_i-\theta_j)\Big],
\end{eqnarray}
where we neglect $\tau$ dependence of $\theta_j$.
We simply assume that $K_j\equiv n_{sj}/(2m_j)=K$, $\Delta_{j0}=\Delta_0$
and $\gamma_{ij}=\gamma_{ji}=\gamma$, namely, all the bands are
equivalent.
Then the action for the phase variables $\theta_j$ is
\begin{equation}
S[\theta]= \frac{\Lambda^{d-2}}{t} \int d^dx\left( \sum_j(\nabla\theta_j)^2
+\alpha\Lambda^2\sum_{i< j}\cos(\theta_i-\theta_j) \right),
\end{equation}
where $t/\Lambda^{d-2}=k_BT/K$ and $\lambda\Lambda^2=2\gamma\Delta_0^2/K$.
We have introduced the cutoff $\Lambda$ so that $t$ and $\alpha$ are
dimensionless parameters.
We assume that $\alpha>0$ in this paper.
We consider the case $N=3$ and discuss the phase transition in this model.
Apparently this model has $S_3$ symmetry.
If we neglect the kinetic term, the ground states is two-fold degenerate.
The two ground states are indexed by the chirality $\kappa$.

We perform a unitary transformation:
$\theta_1 =-2\pi/3-(1/\sqrt{2})\eta_1+(1/\sqrt{6})\eta_2
+(1/\sqrt{3})\eta_3$,
$\theta_2 = -(2/\sqrt{6})\eta_2+(1/\sqrt{3})\eta_3$, and
$\theta_3 = 2\pi/3+(1/\sqrt{2})\eta_1+(1/\sqrt{6})\eta_2
+(2/\sqrt{3})\eta_3$,
where $\eta_i$ ($i=1,2,3$) indicate fluctuation fields.
$\eta_3$ describes the total phase mode,
$\eta_3=(\theta_1+\theta_2+\theta_3)/\sqrt{3}$, and is not important
because this mode turns out to be a plasma mode by coupling with the
long-range Coulomb potential.
The action $S[\eta]\equiv S[\theta]$ becomes
\begin{eqnarray}
S[\eta]&=& \frac{\Lambda^{d-2}}{t}\int d^dx\Big[ \sum_j(\nabla\eta_j)^2
+\alpha\Lambda^2\Bigl( \cos\left( \sqrt{2}\eta_1+\frac{4\pi}{3}\right)\nonumber\\
&+& 2\cos\left( \frac{1}{\sqrt{2}}\eta_1+\frac{2\pi}{3}\right)
\cos\left( \sqrt{\frac{3}{2}}\eta_2\right) \Bigr) \Big].
\end{eqnarray}

\begin{figure}
\begin{center}
  \includegraphics[height=5cm,angle=90]{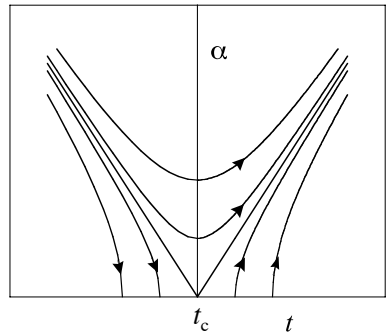}
\caption{Renormalization group flow for the 
sine-Gordon model.
The flow is indicated as $\mu$ is increased ($\mu\rightarrow\infty$).
}
\label{fig:3}       
\end{center}
\end{figure}

This model shows the chiral transition\cite{yan14} as well as the 
Kosterlitz-Thouless transition\cite{kos73}.
The renormalization group method\cite{zin89, ami80} is applied to 
obtain the beta functions.  They are given by
\begin{eqnarray}
\mu\frac{\partial t}{\partial\mu}&=& (d-2)t+At\alpha^2 \\
\mu\frac{\partial \alpha}{\partial\mu}&=& -2\alpha
+\frac{1}{4\pi}\alpha t,
\end{eqnarray}
for the mass parameter $\mu$.  Here $A$ is a constant.
The equation for $\alpha$ has a fixed point at $t=8\pi$.
In two dimension $d=2$ the renormalization group flow is the same
as that for the Kosterlitz-Thouless transition (see Fig.4).

 
There is a chirality transition at a finite temperature where
the states with chirality $\kappa=\pm 1$ disappear and simultaneously
the chirality vanishes.
This is shown by taking account of the fluctuation around the
minimum of the potential.
Using $\cos(\sqrt{3/2}\eta_2)=1-(4/3)\eta_2^2+\cdots$, the action is
written as
\begin{eqnarray}
S&=& \frac{\Lambda^{d-2}}{t}\int d^dx\Big[ \sum_j(\nabla\eta_j)^2
+\alpha\Lambda^2\Bigl( \cos\left(\sqrt{2}\eta_1+\frac{4\pi}{3}\right)\nonumber\\
&-& 2\Bigl|\cos\left( \frac{1}{\sqrt{2}}\eta_1+\frac{2\pi}{3}\right)\Bigr|\Bigr)
\nonumber\\
&+&\frac{3\alpha\Lambda^2}{2}\Bigl|\cos\left( \frac{1}{\sqrt{2}}\eta_1+
\frac{2\pi}{3}\right)\Bigr|\eta_2^2 \Big]. \nonumber\\
\end{eqnarray}
We integrate out the field $\eta_2$ to obtain the effective action.
The effective free-energy density in two dimensions is obtained as
\begin{eqnarray}
\frac{f[\varphi]}{\Lambda^2}&=&  \frac{1}{2}K\Lambda^{-2}(\nabla\varphi)^2
+\epsilon_0\Bigl( \cos\varphi
-2\Bigl|\cos\left(\frac{\varphi}{2}\right)\Bigr|\Bigr)\nonumber\\
&+& \frac{1}{2}k_BT\frac{c}{4\pi}\ln\left(\frac{c\Lambda^d}{t}
+\frac{3\alpha\Lambda^d}{2t}
\Bigl|\cos\left(\frac{\varphi}{2}\right)\Bigr|\right)\nonumber\\
&+& k_BT\frac{3\alpha}{16\pi}
\Bigl| \cos\left(\frac{\varphi}{2}\right)\Bigr|
\ln\Bigl(1+\frac{2c}{3\alpha}
\Bigl| \cos\left(\frac{\varphi}{2}\right)\Bigr|^{-1}\Bigr),\nonumber\\
\end{eqnarray}
for $\varphi\equiv 4\pi/3+\sqrt{2}\eta_1$ where $\Lambda$ is a cutoff,
$c$ is a constant and $\epsilon_0=k_BT\alpha/t=2\gamma\Delta_0^2/\Lambda^2$.
The critical temperature $T_{chiral}$ of the chirality transition
is determined by the condition that we have a minimum at $\varphi=\pi$
(first-order transition).
$T_{chiral}$ is shown as a function of $\alpha$ in Fig.5.
$T_{chiral}=(K/k_B)t_c$ is dependent on $\alpha$, where $\alpha$ is 
proportional to the 
Josephson coupling, while the temperature of the Kosterlitz-Thouless
transition $T_{KT}=(K/k_B)8\pi$ is independent of $\alpha$.
Thus $T_{chiral}$ and $T_{KT}$ are different in general.

We have shown a model which shows a transition due to growing
fluctuations.
The disappearance of the chirality results in the emergency of a
Nambu-Goldstone boson.  This represents the phenomenon that the
Nambu-Goldstone boson appears from a fluctuation effect.
Please note that this does not say that a discreet symmetry
can be broken by Nambu-Goldstone boson proliferation.
A Nambu-Goldstone would emerge as a result of a discree symmetry breaking.
At $T>T_{chiral}$ two spins in Fig.1 are antiferromagnetically 
aligned and one spin vanishes.  This means that the one spin is
rotating freely accompanied with the existence of a massless boson.
Our model shows that the $Z_2$-symmetry breaking induces a massless boson.
If we neglect the kinetic term in the action, $T_{chiral}$ is
determined uniquely as $T_{chiral}=\epsilon_0/2$.
$\epsilon_0$ corresponds to $J$ in the two-dimensional XY model.
This suggests that there is a chirality transition in the 2D XY model
on a two-dimensional triangular lattice at near $T=J/2$,
which has been confirmed by a numerical simulation\cite{miy84}.
The existence of the Kosterlitz-Thouless transition has also been
shown at near $T=J/2$.

We discuss whether our model is applicable to real superconductors.
We expect that our model applies to, for example, layered 
superconductors like cuprates
with small Josephson couplings.
This type of transition has been discussed for three-band
superconductors with frustrated interband Josephson couplings\cite{boj13}.
Recent experiments indicate a possible first-order phase transition
below the superconducting transition temperature in multilayer
cuprate superconductor HgBa$_2$Ca$_4$Cu$_5$O$_y$\cite{tan14}.
We hope that this phase transition is related to the dynamics of
multicomponent order parameters.

\begin{figure}
\begin{center}
  \includegraphics[height=5cm]{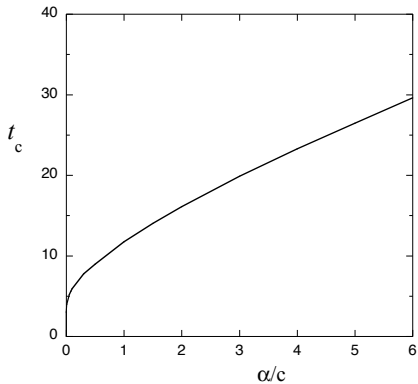}
\caption{$t_c\equiv k_BT_{chiral}/K$ as a function of $\alpha/c$
with $c=4\pi$.
}
\label{fig:4}       
\end{center}
\end{figure}

\section{$SU(N)$ sine-Gordon model}

In this section let us consider a generalized Josephson interaction
where the Josephson term is given by a $G$-valued sine-Gordon potential
for a compact Lie group $G$.
This model includes multiple excitation modes, and is
a nonabelian generalization of the sine-Gordon model.
The Lagrangian is written as
\begin{equation}
\mathcal{L}= \frac{1}{2t}{\rm Tr}\partial_{\mu}g\partial^{\mu}g^{-1}
+\frac{\alpha}{2t}{\rm Tr}(g+g^{-1}),
\end{equation}
for $g\in G$.  When $g=e^{i\varphi}\in U(1)$, this Lagrangian is
reduced to that of the conventional sine-Gordon model.
This model can be regarded as the chiral model with the mass term.
Here we consider the $SU(N)$ or $O(N)$ model: $G=SU(N)$ or $O(N)$.
In the limit $t\rightarrow\infty$ with keeping $\lambda\equiv \alpha/t$ 
constant, the
$SU(N)$ sine-Gordon model is reduced to a unitary matrix model.
It has been shown by Gross and Witten that, in the large $N$ limit 
with the coupling
constant $\lambda=N\beta$, for the model $N\beta{\rm Tr}(g+g^{\dag})$, 
there is a third-order
transition at some critical $t_c$\cite{gro80}.
Brezin and Gross considered the model to generalize the coupling constant
$\lambda$ to be a matrix and also found that there is a phase 
transition\cite{bre80, bro81, bre10}.
Recently, the vortex structure for a nonabelian sine-Gordon model was
investigated numerically\cite{nit15}. 

An element $g\in G$ is represented in the form:
\begin{equation}
g= g_0\exp\left(i\lambda \sum_aT_a\pi_a\right),
\end{equation}
where $\lambda$ is a real number $\lambda\in {\bf R}$ and $g_0\in G$ is
a some element in $G$.
We put $g_0=1$ in this paper.
$T_a$ ($a=1,2,\cdots, N_T$) form a basis of the Lie algebra of $G$.
$N_T=N^2-1$ for $SU(N)$ and $N_T=N(N-1)/2$ for $O(N)$.
$\{T_a\}$ are normalized as
\begin{equation}
{\rm Tr}T_aT_b= c\delta_{ab},
\end{equation}
with a real constant $c$.
The scalar fields $\pi_a$ indicate fluctuations around the classical 
solution, that is, the nonabelian perturbation to the state $g_0$.
We expand $g$ by means of $\pi_a$ as
\begin{equation}
g= g_0\Big[ 1+i\lambda T_a\pi_a-\frac{1}{2}\lambda^2(T_a\pi_a)^2
+\cdots \Big],
\end{equation}
and evaluate the beta functions of renormalization group theory.

The renormalization group equations read\cite{yan15}
\begin{eqnarray}
\mu\frac{\partial t}{\partial\mu}&=& (d-2)t-\frac{C_2(G)}{4}t^2
+A_0C(N)t\alpha^2,\\
\mu\frac{\partial \alpha}{\partial\mu}&=& -\alpha\left( 2-C(N)t\right),
\end{eqnarray}
where $A_0=A_0(N)$ is a constant (depending on $N$), $c=1/2$ and the volume 
element $\Omega_d/(2\pi)^d$
is included in the definition of $t$ for simplicity.
$C(N)$ is the Casimir invariant in the fundamental representation
given by
\begin{eqnarray}
C(N)&=& c\frac{N^2-1}{N}~~{\rm for}~~G=SU(N),\\
    &=& c\frac{N-1}{2} ~~{\rm for}~~G=0(N).
\end{eqnarray}
The coefficient of $t^2$ term in 
$\mu\partial t/\partial\mu$ is the Casimir invariant in the adjoint
representation defined by $\sum_{ab}f_{abc}f_{abd}=C_2(G)\delta_{cd}$.
$C_2(G)$ for $SU(N)$ is given as
\begin{equation}
C_2(G)=2Nc~~ {\rm for}~~ G=SU(N).
\end{equation}
$C_2(G)$ for $G=O(N)$ is proportional to $(N-2)c$.
Thus beta functions are determined by Casimir invariants.
 
There is a zero of beta functions in two dimensions($d=2$):
\begin{equation}
t_c=\frac{2}{C(N)},~~\alpha_c= \sqrt{\frac{C_2(G)}{2A_0(N)}}\frac{1}{C(N)}.
\end{equation}
This is a bifurcation point that divides the parameter space
into two regions.  One is the strong coupling region where
$\alpha\rightarrow\infty$ as $\mu\rightarrow\infty$, and the other
is the weak coupling region where $\alpha\rightarrow 0$ as 
$\mu\rightarrow\infty$.
In the weak coupling region, we can use a perturbation theory by
expanding $g$ by means of the fluctuation fields $\pi_a$.
This results in the existence of multiple frequency modes.
We expect that these modes may be observed.
There may be a possibility to classify excitation modes using a
group theory.

\section{Summary}

The Nambu-Goldstone-Leggett modes play a nontrivial role
in multi-condensate superconductors.
We discussed the plasma and Leggett modes, 
time-reversal symmetry breaking, half-quantized flux
vortex and its structure as a monopole in a superconductor, 
massless modes, and generalized sine-Gordon models. 
An $N$-gap superconductor has $N-1$ phase-difference variables, and
the $U(1)^{N-1}$ phase invariance can be partially or totally broken.  
The $N-1$ phase modes are in general massive due to the symmetry breaking
by Josephson couplings.
When the Josephson couplings are frustrated, symmetry is partially
broken and some of phase modes can become massless.

When the phase fluctuation is large, we cannot use a perturbation
in the phase variable $\varphi$ as done by Leggett. 
In this case a kink solution provides a
new excitation mode.
A fractional-quantum-flux vortex exists at the edge of the kink.
An effect of fluctuation is investigated, based on a toy model,
where fluctuation restore time reversal symmetry from the ground
state with time-reversal symmetry breaking.
We also proposed a $G$-valued sine-Gordon model as a generalization
of the sine-Gordon model.
If we neglect spatial dependence of $G$-valued fields, this model is
reduced to a unitary matrix model.
We derived a set of renormalization group equations for this model.
In the weak coupling region the perturbative procedure may lead to
multiple excitation modes.

\section{acknowledgments}
The author expresses his sincere thanks to J. Kondo, K. Yamaji, I. Hase
and Y. Tanaka for helpful discussions.
He also expresses his sincere thanks to Prof. S. Hikami at the Okinawa
Institute for Science and Technology for stimulating discussions.


\end{document}